\documentstyle[12pt]{article}

\newcommand{\plabel}{\label}
\newcommand{\cD}{{\cal D}}
\newcommand{\oB}{\vert_{\partial M}=0}
\newcommand{\cB}{{\cal B}}
\begin{document}
\title{The Faddeev--Popov trick in the presence of boundaries}
\author{Dmitri V.~Vassilevich}
\date{ }
\maketitle
\begin{center}{\it Department of Theoretical Physics,
St.Petersburg University}\\{\it 198904 St.Petersburg,
Russia}\end{center}
\begin{abstract}
We formulate criteria of applicability of the Faddeev--Popov
trick to gauge theories on manifolds with boundaries. With
the example of Euclidean Maxwell theory we demonstrate that
the path integral is indeed gauge independent when these criteria
are satisfied, and depends on a gauge choice whenever these
criteria are violated.
\end{abstract}
\newpage
\section{Introduction}
Modern interest to quantum field theory on manifolds with boundary
is motivated by applications to quantum cosmology and the Casimir
effect. Over last few years, substantial progress has been made
in calculation of the heat kernel asymptotics and functional
determinants. However, the situation with gauge field contribution
is far from being clear (for a review, see recent monograph
\cite{EsKaP}). One of the most important problems
is gauge dependence of on--shell effective action \cite{EsKaP}.
The simplest way to demonstrate gauge independence of the
path integral is given by the Paddeev-Popov trick \cite{FP}.
The aim of the present paper is to formulate criteria of
applicability of this trick on manifolds with boundary.
Two gauge conditions give equivalent path integrals if they
are admissible for the same set of gauge invariant boundary
conditions. Admissibility means that a gauge condition eliminates
all linearized gauge transformations in a unique way. Gauge
invariance may be replaced by BRST invariance of boundary conditions
\cite{MS}. The criteria are formulated in the next section.
In the Section 3 we consider an example of Euclidean Maxwell
theory. We show that the path integral is gauge independent
when the criteria of Sec. 2 are satisfied. For generic choice
of boundary geometry the path integral becomes gauge dependent
whenever these criteria are violated. This means that to achieve
admissibility one should choose gauge dependent boundary conditions.
Such boundary conditions describe different physics. In the Appendix
we collect all geometric notations and expressions for the
heat kernel asymptotics.
\section{General gauge theories}
Consider a gauge theory with classical action $S(\Phi )$ being
invariant under infinitesimal gauge transformations
$\delta_\xi \Phi =G\xi$. The path integral is given by the
expression:
\begin{equation}
Z(\alpha , \chi )=\int \cD \Phi J(\chi )
\exp (-S(\Phi ) -\frac 1{2\alpha} \chi^2) ,
\plabel{eq:gf}
\end{equation}
where $\chi$ is a gauge fixing condition, $J(\chi )$ is the
Faddeev--Popov determinant, $J=\det (-\chi G)$.
We assume that $Z$ depends on external geometry of
the space-time domain and on boundary conditions for
the quantum field $\Phi$. We do not introduce any
sources or background fields explicitly.
If background field corresponding to the quantum filed
$\Phi$ is present, we must assume that the background
is taken "on-shell", i.e. satisfying equations of motion.
It is well known that the path integral
(\ref{eq:gf}) can be obtained from another path integral
\begin{equation}
Z(\chi_a ) =\int \cD \Phi J(\chi ) \delta (\chi -a)
\exp (-S(\Phi )) \plabel{eq:ca}
\end{equation}
after averaging over $a$ with the weight $\exp
(-\frac 1{2\alpha} a^2)$.
Hence,
it is enough to study gauge-independence of the path
integral
\begin{equation}
Z(\chi ) =\int \cD \Phi J(\chi ) \delta (\chi )
\exp (-S(\Phi )) \ . \plabel{eq:c}
\end{equation}
The equivalence of two path integrals,
$Z(\chi_1 )$ and $Z(\chi_2 )$, can be established by using the
Faddeev-Popov trick. One should use twice the following
representation of unity
\begin{equation}
1=\int \cD \xi J(\chi ) \delta (\chi (\Phi +G\xi )),
\plabel{eq:un}
\end{equation}
One should insert (\ref{eq:un}) with
$\chi =\chi_2$ in the integrand of $Z(\chi_1)$, change
integration variables to $\Phi -\delta_\xi \Phi$, and use
again eq. (\ref{eq:un}) with $\chi =\chi_1$. This procedure
can be done successfully if the two gauges $\chi_1$ and
$\chi_2$ satisfy the following requirements.

    (i) {\it Gauge-invariance of the boundary conditions}.
Let
\begin{equation}
\cB \Phi \oB \plabel{eq:bp}
\end{equation}
be a boundary condition for the fields $\Phi$ with some
boundary operator $\cB$. There should exist boundary
conditions
\begin{equation}
{\cB}_\xi \xi \oB \plabel{eq:bx}
\end{equation}
for gauge transformation parameters $\xi$ such
that
\begin{equation}
\cB \delta_\xi \Phi \oB \ . \plabel{eq:inv}
\end{equation}

    The eq. (\ref{eq:inv}) means that gauge transformations
map the functional space defined by eq. (\ref{eq:bp}) onto
itself for some boundary conditions (\ref{eq:bx}) imposed
on gauge parameter $\xi$. It is clear that the operator
$\cB_\xi$ defines boundary conditions for the ghost fields.

We use twice the integral (\ref{eq:un}) over the same functional
space. Hence, the operators $\cB_\xi$ are to be the same for
both gauges $\chi_1$ and $\chi_2$.

(ii) {\it Admissibility of $\chi_1$ and $\chi_2$}.
 We call a gauge condition $\chi$ admissible
if for given gauge-invariant boundary conditions
(\ref{eq:bp}), (\ref{eq:bx}) the equation
\begin{equation}
\chi (\Phi +G\xi )=0 \plabel{eq:adm}
\end{equation}
has unique solution{\footnote{Note that we consider
only linearized gauge transformations thus
avoiding the question of Gribov ambiguities.
This restriction is correct at least at the one-loop
approximation}}
 $\xi$ for every $\Phi$. Again, both gauges $\chi_1$ and
$\chi_2$ should be admissible for the same
boundary operators $\cB$ and $\cB_\xi$.

If path integral in one gauge can not be transformed to
another gauge by the Faddeev--Popov trick, they most probably
describe different physics. More precisely, such gauges
require different boundary conditions for their selfconsistent
formulation. These boundary conditions may describe different
physics.

\section{Two examples}
Consider the action for Maxwell field
on $m$-dimensional Euclidean manifold $M$:
\begin{equation}
S=\int_M d^mx \sqrt g \left ( \frac 14 F_{\mu\nu}F^{\mu\nu}
+\frac 1{2} [\chi (A)]^2 \right )
\plabel{act}
\end{equation}
Suppose for simplcicty that the metric $g_{\mu\nu}$ is flat.
We shall compare different gauge conditions to the Lorentz gauge
\begin{equation}
\chi_L=\nabla^\mu A_\mu . \plabel{Lg}
\end{equation}
Ghost operator takes the form of ordinary Laplacian,
$\chi_L(\nabla \xi )=\Delta \xi$. Note, that constant ghosts
should be excluded (see, e.g. \cite{dprd}). Near the boundary
the gauge fixing function is $\chi =(\nabla_m -L_{ii})A_m+
\tilde \nabla^iA_i$, where subscript $m$ denotes normal component
of a vector, $\nabla$ is covariant derivative on $M$,
 $L_{ii}$ is trace of the second fundamental form
on the boundary, $\tilde \nabla$ is covariant derivative on the
boundary.

Let us choose the so called relative boundary conditions for $A_\mu$
and Dirichlet boundary conditions for the ghosts:
\begin{equation}
(\nabla_m-L_{ii})A_m \oB ,\quad A_i\oB ,\quad \xi \oB .
\plabel{rbc}
\end{equation}
Gauge invariance of the boundary conditions (\ref{rbc}) is equivalent
to the equation $\Delta \xi\oB$, which is obvious for eigenfunctions
of the ghost operator $\Delta$. The equation $\nabla^\mu A_\mu =
-\Delta \xi$ has unique solutions for every $A_\mu$ because
$\nabla^\mu A_\mu \oB$, and $\Delta $ is invertible on the
space of Dirichlet fields without constant zero mode. Hence the
gauge (\ref{Lg}) with boundary conditions (\ref{rbc}) is admissible.

The path integral takes the form:
\begin{equation}
Z_L={\det}_V(-\Delta )^{-\frac 12}{\det}_S(-\Delta ),
\plabel{Lpi}
\end{equation}
where the first determinant is taken over vector fields and
the second one is calculated for scalars.

It was demonstrated in \cite{dprd} that the path integral
(\ref{Lpi}) is equivalent to the Hamiltonian path integral
with covariant path integral measure.

\subsection{An admissible gauge}
Let the manifold $M$ admits a metric such that $g_{00}=1$,
$g_{0i}=0$. Let the boundaries correspond to $x^0=const$
surfaces. Consider the gauge
\begin{equation}
\chi =\alpha^{-\frac 12} (\nabla^\mu A_\mu +f(x^0)\tilde\nabla^i
A_i ) \plabel{g31}
\end{equation}
where $\tilde\nabla$ is covariant derivative on $m-1$-dimensional
slices, $\alpha$ is a constant, $f(x^0)$ is an arbitrary function,
$f>-1$.

It is easy to see that the  gauge (\ref{g31}) supplemented by the
relative boundary conditions (\ref{rbc}) satisfies both requirements
(i) and (ii) of the previous section. Indeed, (i) was already
demonstrated above. Equation (ii) gives
\begin{equation}
\chi (A)+L_\chi \xi =0, \quad L_\chi =\alpha^{-\frac 12}
(\Delta +f(x^0)\tilde\Delta ) ,
\plabel{ii31}
\end{equation}
$\tilde\Delta =\tilde \nabla^2$.
One can check that both $\chi (A)$ and $L_\chi \xi$ vanish on the
boundary if $A$ and $\xi$ satisfy (\ref{rbc}).
$L_\chi$ is self-adjoint
(or at least symmetric) operator. Hence, if we neglect possible
topological obstructions, (\ref{ii31}) has unique solution $\xi$
for any $A$.

Consider the path integral
\begin{equation}
Z=\int \cD A_\mu \det (-L_\chi ) \exp \left ( -\int d^4x \sqrt{g}
(\frac 14 F_{\mu\nu}F^{\mu\nu} +\frac 1{2} \chi^2 ) \right )
\plabel{p31}
\end{equation}
Let us change variables in (\ref{p31}):
\begin{eqnarray}
A_\mu &=&A_\mu^T +\partial_\mu \phi ,\quad
\nabla^\mu A_\mu^T=0 \nonumber \\
\cD A_\mu &=& \cD A_\mu^T \cD \phi {\det}_S^{\frac 12} (-\Delta )
\plabel{c31}
\end{eqnarray}
This change is consistent with boundary conditions in question.
Mixing between $\phi$ and $A^T$ can be removed by a shift of
$\phi$, $\phi \to \phi' =\phi +\alpha^{\frac 12}
L_\chi^{-1}\tilde\nabla^iA_i^T$, which does not change
boundary conditions for $\phi$ and produces unit Jacobian factor.
Integration over $\phi'$ is immediately performed giving
${\det}(-L_\chi )^{-1}$. As before, integration over $A^T$ gives
${\det}_T(-\Delta )^{-\frac 12}$. Collecting all contributions
together, we arrive at the path integral
\begin{equation}
Z={\det}_T(-\Delta )^{-\frac 12} {\det}_S^{\frac 12} (-\Delta )
\end{equation}
This coincides with the result (\ref{Lpi}) in the Lorentz gauge
after taking into account factorization property of the vector
determinant ${\det}_V(-\Delta )={\det}_T(-\Delta){\det}_S(-\Delta)$
which holds for relative boundary conditions.

There are two important particular cases of the gauge condition
(\ref{g31}). $f(x^0)\to \infty$ corresponds to the Coulomb gauge,
while $f(x^0)\to -1$ gives axial gauge. On manifolds with boundaries
such gauges were studied in \cite{EK} for the case of Euclidean
Maxwell theory and in \cite{AEK} for quantum gravity. For
$f=1$ (or $f=\infty$) the ghost operator $L_\chi$ is not elliptic.
Spectrum of $L_\chi$ becomes infinitely degenerate and the heat
kernel technique is not applicable. A more careful way to introduce
such gauges is to consider a limiting procedure from (\ref{g31}).

\subsection{Esposito gauges}
Consider the gauge fixing condition depending on an arbitrary
vector field $B^\mu$:
\begin{equation}
\chi_E=(\nabla^\mu +B^\mu )A_\mu
\plabel{chiE}
\end{equation}
Suppose that on a boundary $B^\mu$ is parallel to normal vector
$e_m$, $B^a\vert_{\partial M}=B(x)\delta^a_m$. $a,b,c$ will denote
flat tangential indices on $M$, $B^a=e^a_\mu B^\mu$. Near the
boundary the gauge (\ref{chiE}) reads: $\chi_E=(\nabla_m-L_{ii}
+B)A_m+\tilde \nabla^iA_i$.

The boundary conditions
\begin{equation}
(\nabla_m-L_{ii}+B)A_m \oB ,\quad A_i\oB ,\quad \xi \oB
\plabel{Ebc}
\end{equation}
ensure admissibility of the gauge (\ref{chiE}) in the sense of
the previous section. These boundary conditions depend on $B$.
Hence the Faddeev--Popov trick can not be used to demonstrate
gauge independence of the path integral. One could choose
relative boundary conditions which do not depend on $B$.
In this case, however, the gauge (\ref{chiE}) will no
longer be admissible.

The gauge (\ref{chiE}) generalises a family of gauges considered
by Esposito, Kamenshchik and co-workers \cite{EKMP} on manifolds
with spherical boundaries. Namely, these authors calculated
the one--loop conformal anomaly ${\cal A}$ on a ball ("one--boundary
problem") and in a region between two concentric spheres
("two--boundary problem") for $m=4$ and $B=const.\times \frac 1r$,
where $r$ is radial coordinate. They found that ${\cal A}$ depends
on $B$ for the one--boundary problem and is $B$-independent for
the two--boundary problem.  According to the authors \cite{EKMP},
gauge dependence in the former case is due to a singularity of the
$3+1$ decomposition at $r=0$. According to the present author
\cite{qg6}, gauge independence in the latter case is totally
due to special choice of geometry which allows for cancellation
of contribution of the two boundaries.

Since integration by part does not introduce any
surface terms, we can represent (\ref{act}) in the following
form
\begin{equation}
S=-\frac 12 \int d^mx \sqrt g A^a (D^\mu D_\mu +E)_a^bA_b
\plabel{act1}
\end{equation}
where $a,b$ are flat tangential indices, $A^a=A^\mu e_\mu^a$.
New covariant derivative $D_\mu =\nabla_\mu +\omega_\mu$ contains
an auxiliary connection field
\begin{equation}
\omega_\mu^{ab}=\frac 12 (e^a_\mu B^b -e^b_\mu B^a)
\plabel{ome}
\end{equation}
The matrix $E$ has the form
\begin{equation}
E_{ab}=\frac 12 (\nabla_aB_b+\nabla_bB_a)
+\frac 14 (\delta_{ab}B^2 +(m-6)B_aB_b)
\plabel{E}
\end{equation}
The ghost operator corresponding to the gauge fixing
term (\ref{chiE}) is
\begin{equation}
L^{gh}=-(\nabla^\mu \nabla_\mu +B^\mu \nabla_\mu )
\plabel{Lgh}
\end{equation}
First order derivative term can be removed again by introducing
a new connection:
\begin{equation}
L^{gh}=-(D^{[gh]\mu}D^{[gh]}_\mu +E^{[gh]}), \quad
\omega_\mu^{[gh]} =\frac 12 B_\mu , \quad
E^{[gh]}=-\frac 12 \nabla^\mu B_\mu -\frac 14 B^2
\plabel{Eog}
\end{equation}

The path integral is given by a product of two determinants:
\begin{equation}
Z_E=\det (-(D^\mu D_\mu +E))^{-\frac 12} \det (L^{gh})
\plabel{dets}
\end{equation}
To study the problem of gauge dependence of $Z_E$ let us
use gauge--invariant zeta-function regularization and
evaluate scaling behaviour (conformal anomaly), which is
given by
\begin{equation}
{\cal A} =\zeta_{ph}(0)-2\zeta_{gh} (0)
\plabel{cA}
\end{equation}
where two terms represent individual contributions of the
photon and ghost operators in (\ref{dets}). Right hand side
of (\ref{cA}) can be calculated by using the heat kernel
expansion and the relation $\zeta_L(0)=a_m (L)$. Using
expressions for $a_m$ from the Appendix, one can calculate
${\cal A}$ for $m=2,3,4$ and arbitrary $B(x)$ and boundary
geometry. We observe the following properties:

\begin{enumerate}
\item
Gauge dependence in the volume integrals is cancelled.
\item
For generic boundary geometry the boundary terms are gauge
dependent ($m=3,4$).
\item
For the two--boundary problem and $B=const.\times \frac 1r$
contributions of the two boundaries cancel each other.
\item
For the one--boundary problem $E$ and $E^{[gh]}$ are singular
at $r=0$. Individual contributions of ghosts and photons can
not be calculated by using formulas from the Appendix.

\end{enumerate}
Both explanations \cite{EKMP,qg6} to gauge dependence of the
conformal anomaly in Esposito gauge are true. One--boundary
case really contains a dangerous singularity. Gauge independence
in two--boundary problem is really due to a very special
choice of geometry. In general, Esposito gauge gives
gauge dependence of the conformal anomaly in complete agreement
with statements of Sec.~2.

Though the problem of gauge dependence has received an explanation
from the mathematical point of view, physical consequences are not
clear. In Lorentzian signature of space-time both relative and
Esposito boundary conditions correspond to electromagnetic field
in a conducting cavity. However, some details of interaction of
photons with material of a boundary must be changed. A useful
test would be to evaluate vacuum expectation value of $J^iJ_i$
with boundary values of the current $J^\mu=\nabla_\nu F^{\nu\mu}$.
This can be done, in principle, after extension of the results
\cite{BGV2} to mixed boundary conditons.
\section{Conclusions}
In the present paper we formulated some simple criteria of
applicability of the Faddeev--Popov trick on manifolds with
boundaries. Namely, if two gauges are admissible for the same
set of gauge invariant boundary conditions imposed on ghosts
and gauge fields, they give identical path integrals. As an
example, Euclidean Maxwell theory was considered. We demonstrated
that for the family of gauges (\ref{g31}) the above criteria
are satisfied and the path integral is indeed gauge independent.
Violation of these criteria for the Esposito gauges leads to
gauge dependence of the path integral. Physical consequences
of this effect are still to be clarified.
\section*{Acknowledgments}
The author is grateful to Ivan Avramidi, Giampiero
Esposito and Alexander Kamentshchik for discussions,
and especially to Andrei Barvinsky for valuable
comments. This work was supported by the Russian
Foundation for Fundamental Research, grant 97-01-01186.
\section*{Appendix: The heat kernel coefficients}
In this appendix we give general expressions for the
heat kernel asymptotics with mixed boundary conditions
\cite{BG,jmp1}. Let $M$ be a compact smooth
manifold of dimension $m$ with smooth boundary $\partial M$.
Let $L$ be an operator of Laplace type on the space of
smooth sections $C^\infty (V)$ of certain vector
bundle over $M$. This means that by
introducing suitable metric and connection fields it can be
represented as
\begin{equation}
L=-(g^{\mu\nu}D_\mu D_\nu +E) \plabel{Lap}
\end{equation}
where $E$ is an endomorphism.

We must impose suitable boundary conditions. Let $\Phi \in
C^\infty (V)$. Dirichlet boundary conditions are
\begin{equation}
{\cal B}\Phi =\Phi \oB \plabel{Diri}
\end{equation}
Choose an orthonormal frame on $M$ such that $e_m$ is
inward pointing unit vector, $\{ e_i\}$ is orthonormal frame
on $\partial M$. Let $S$ be an endomorphism of $V$ defined on
$\partial M$. Neumann boundary conditions are
\begin{equation}
{\cal B}\Phi =(D_m +S)\Phi \oB
\plabel{Neu}
\end{equation}
One can also introduce mixed boundary conditions. We assume given
a decomposition $V=V_N\oplus V_D$ near $\partial M$. We take Neumann
boundary conditions on $V_N$ and Dirichlet boundary conditions on
$V_D$. $S$ acts only on $V_N$ and zero on $V_D$. Let $\Pi_N$ and
$\Pi_D$ be the corresponding projection operators and let
$\psi =\Pi_N -\Pi_D$. Such boundary conditions are elliptic.

As $t\to +0$, there is an asymptotic expansion
\begin{equation}
{\rm Tr}_{L^2} (e^{-tL})\sim \sum_{n=0}^\infty
a_n(D,{\cal B}) t^{(n-m)/2} \plabel{asym}
\end{equation}
where the coefficients depend on the boundary operator ${\cal B}$.
Suppose that the metric $g$ is flat. Then
\begin{eqnarray}
a_0&=&(4\pi )^{-m/2} {\rm tr} ({\bf 1})[M] \nonumber \\
a_1&=&(4\pi )^{-(m-1)/2}\frac 14 {\rm tr}(\psi )[\partial M]
\nonumber \\
a_2&=&(4\pi )^{-m/2}\frac 16 {\rm tr}\{ 6E[M]+
(2L_{ii}+12S) [\partial M]\} \nonumber \\
a_3&=&(4\pi )^{-(m-1)/2}\frac 1{384} {\rm tr} \{
96\psi E +(13\Pi_N-7\Pi_D)L_{ii}L_{jj} \nonumber \\
\ &\ &+(2\Pi_N+10\Pi_D )L_{ij}L_{ij}+96SL_{jj}+192S^2
-12\psi_{:i}\psi_{:i} \} [\partial M] \nonumber \\
a_4&=&(4\pi )^{-m/2}\frac 1{360} {\rm tr}\{
(60E_{;\mu\mu}+180E^2+30\Omega_{\mu\nu}\Omega^{\mu\nu}) [M]
\nonumber \\
\ &\ &+((240\Pi_N-120\Pi_D)E_{;m} +120EL_{ii}+\frac 1{21}
((280\Pi_N+40\Pi_D)L_{ii}L_{jj}L_{kk} \nonumber \\
\ &\ &+(168\Pi_N-264\Pi_D)L_{ij}L_{ij}L_{kk}
+(224\Pi_N+320\Pi_D )L_{ij}L_{jk}L_{ki} ) \nonumber \\
\ &\ &+720SE+144SL_{ii}L_{jj} +48SL_{ij}L_{ij}+480S^2L_{ii}
+480S^3 \nonumber \\
\ &\ &+60\psi \psi_{:i}\Omega_{im}-12\psi_{:i}\psi_{:i}L_{jj}
-24\psi_{:i}\psi_{:j}L_{ij}-120\psi_{:i}\psi_{:i}S ) [\partial M]\}
\end{eqnarray}
Here ";" and ":" denote covariant differentiation on $M$ and
$\partial M$ respectively. Note, that $E_{:ij}$ and $E_{;ij}$ do
not coincide. Their difference is proportional to the second
fundamental form of the boundary $L_{ij}$. $\Omega_{\mu\nu}=
[D_\mu ,D_\nu ]$. $[M]$ and $[\partial M]$ denote integration
over $M$ and $\partial M$ respectively with proper volume
elements.

\end{document}